\documentclass[aps,prl,floats,twocolumn,showpacs,superscriptaddress]{revtex4}

\usepackage{epsfig}
\usepackage{latexsym,amsmath}

\begin{document}
\title{Competition and adaptation in an Internet evolution model}

\author{M$^{\mbox{\underline{a}}}$ \'Angeles Serrano, Mari{\'a}n Bogu{\~n}{\'a}
and Albert D\'{\i}az-Guilera}

\affiliation{Departament de F{\'\i}sica Fonamental, Universitat de
  Barcelona, Av. Diagonal 647, 08028 Barcelona, Spain}

\date{\today}

\begin{abstract}
We model the evolution of the Internet at the Autonomous System
level as a process of competition for users and adaptation of
bandwidth capability. We find the exponent of the degree
distribution as a simple function of the growth rates of the
number of autonomous systems and the total number of connections
in the Internet, both empirically measurable quantities. This fact
place our model apart from others in which this exponent depends
on parameters that need to be adjusted in a model dependent way.
Our approach also accounts for a high level of clustering as well
as degree-degree correlations, both with the same hierarchical
structure present in the real Internet. Further, it also highlights the interplay between bandwidth, connectivity and traffic of the network.
\end{abstract}


\maketitle

A statistical physics approach to Internet modeling will be
successful only if its large-scale properties can be explained and
predicted on the basis of the interactions between basic units at
the microscopic level \cite{Romusbook,Mendesbook}. Dynamical
evolution rules acting at the local scale would then determine the
behavior and the emergent structural properties of the whole
Internet, which self-organizes under an absolute lack of
centralized control \cite{Barabasi02,Dorogovtsev02}. This approach
is at the core of a set of recent network models focusing on
evolution, which recognise growth as one of the key mechanisms on
network formation, along with preferential attachment or other
utility rules \cite{Barabasi99,Huberman,Goh,Capocci,Fayed,Zhou}. While
several of such models succeed in depicting some of the Internet
features, none of them accounts for a complete description of the
real topology \cite{Falou99,Alexei,Alexei02}. In this paper, we
present a new growing network model which, from competition and
adaptation mechanisms, reproduces the topological properties
observed in the autonomous system level maps of the Internet,
namely: i) a scale-free distribution of the number of connections --or degree-- of vertices $k_i$, characterized by a power law $P(k)\sim k^{-\gamma}$, $2.1 \leq \gamma <2.5$, ii) high
clustering coefficient $c_k$, defined as the ratio between the
number of connected neighbors of a node of degree $k$ and the
maximum possible value averaged for all nodes of degree $k$, and,
finally, iii) disassortative degree-degree correlations, quantified by means of
the average nearest neighbors degree of nodes of degree $k$,
$\bar{k}_{nn}(k)$ \cite{Alexei}.

\begin{figure}
\epsfig{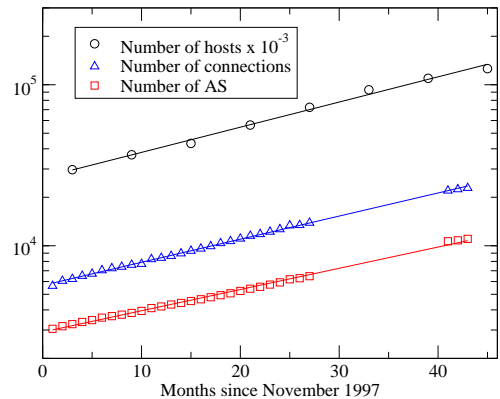} \caption{Temporal
evolution of the number of hosts, autonomous systems and
connections among them from November 1997 to May 2002. Solid lines
are the best fit estimates which gives the values for the rate
growths of $\alpha=0.036$, $\beta=0.029$ and $\delta=0.033$ (units
are month$^{-1}$).}
 \label{evolution}
\end{figure}

We start our analysis by looking at the growth of the Internet
during the last three decades. We focus on the temporal evolution of the number of hosts present in the Internet
\cite{Hosts} as compared to
the number of distinct autonomous systems (ASs) and the total
number of connections among them. We have reanalysed AS maps
collected by the {\it Oregon route-views} project which has
recorded the Internet topology at the AS level since November 1997 \cite{Evolution}. Let $W(t)$, $N(t)$ and $E(t)$ be the total number of hosts (we assume that number of hosts is equivalent to number of users), number of ASs and edges among ASs at
time $t$ respectively. Fig.\ref{evolution} shows empirical
measurements for these quantities revealing exponential
growths, $W(t)\sim W_0 e^{\alpha t}$, $N(t)\sim N_0
e^{\beta t}$ and $E(t)\sim E_0 e^{\delta t}$ respectively,  with rates $\alpha=0.036$, $\beta=0.029$ and $\delta=0.033$, where $\alpha \gtrsim
\delta \gtrsim \beta$. These exponential growths, in turn,
determine the scaling relations with the system size, that is, $W
\sim N^{\alpha/\beta}$, $E \sim N^{\delta/\beta}$ and $\langle k
\rangle \sim N^{\delta/\beta-1}$ \cite{Dorogovtsev01}. All three rates are, indeed,
quite close to each other. This result poses the question of
whether these inequalities actually hold or, in contrast, are due
to statistical fluctuations. A simple argument will convince us
that the inequalities are, actually, the natural answer. There are two mechanisms capable to compensate an increase in the number of users: the creation of new ASs and the creation of new connections by old ASs. When both mechanisms take place
simultaneously, the rate of growth of new ASs, $\beta$,  must
necessarily be smaller than $\alpha$, whereas the rate of growth of
the number of connections, $\delta$, must be greater than $\beta$.
Any other situation would lead to an imbalance between the number
of users and the maximum number of users that the system can
manage. 

Our model is defined according to the following rules: (i) At rate $\alpha W(t)$, new users joint the system and choose provider $i$ according to some preference function, $\Pi_i(\{\omega_j(t)\})$, where
$\omega_j(t)$, $j=1,\cdots,N(t)$, is the number of hosts connected to AS $j$ at time $t$. The function $\Pi_i(\{\omega_j(t)\})$
is normalised so that $\sum_i \Pi_i(\{\omega_j(t)\})=1$ at any time. (ii) At rate $\beta
N(t)$, new ASs join the network with an initial number of users,
$\omega_0$, randomly withdrawn from the pool of users already
attached to existing ASs. Therefore, $\omega_0$ can
be understood as the minimum number of users required to keep ASs in business. (iii) At rate $\lambda$, each user changes his provider and chooses a new one using the same preference function  $\Pi_i(\{\omega_j(t)\})$. Finally, (iv) each node tries to adapt its number of connections to other nodes according to its present number of users, in an attempt to provide them an adequate access to the Internet. We will discuss this last point in the second part of the work. With the above ingredients, in the continuum approximation, the dynamics of single nodes is described by the stochastic
differential equation
\begin{equation}
\frac{d \omega_i}{dt}=A(\omega_i,t)+\left[D(\omega_i,t)\right]^{1/2} \xi(t),
\label{langevin}
\end{equation}
where $A(\omega_i,t)$ is a time dependent drift given by
\begin{equation}
A(\omega_i,t)=(\alpha+\lambda) W(t) \Pi_i-\lambda \omega_i-\beta \omega_0
\label{drift}
\end{equation}
and the diffusion term by
\begin{equation}
D(\omega_i,t)=(\alpha+\lambda) W(t) \Pi_i+\lambda
\omega_i+\beta \omega_0-2\lambda \omega_i \Pi_i.
\label{diffusion}
\end{equation}
Application of the Central Limit Theorem guaranties the
convergence of the noise $\xi(t)$ to a gaussian white noise
in the limit $W(t) \gg 1$. The first term in the right hand side
in Eq.~(\ref{drift}) is a creation term accounting for new and old
users that choose node $i$ as a provider. The second term
represent those users who decide to change their providers and,
finally, the last term corresponds to the decrease of users due to
introduction of newly created ASs. To proceed further, we need to
specify the preference function $\Pi_i(\{\omega_j(t)\})$. We assume that, as a result of a competition process, bigger ASs get users more easily than small ones. The simplest function satisfying this condition corresponds to
the linear preference, that is,
\begin{equation}
\Pi_i(\{\omega_j(t)\})=\frac{\omega_i}{W(t)}
\label{preferential}
\end{equation}
where $W(t)=\omega_0 N_0 \exp{(\alpha t)}$.  In this case, the
stochastic differential equation (\ref{langevin}) reads
\begin{equation}
\frac{d \omega_i}{dt}=\alpha \omega_i-\beta
\omega_0+\left[(\alpha+2\lambda)\omega_i+\beta \omega_0
\right]^{1/2} \xi(t). \label{langevin2}
\end{equation}
Notice that reallocation of users ({\it i.e.} the $\lambda$-term) only
increases the diffusive part in Eq.~(\ref{langevin2}) but has no
net effect in the drift term, which is, eventually, the leading term. The
complete solution of this problem requires to solve the
Fokker-Plack equation corresponding to Eq.~(\ref{langevin2}) with a
reflecting boundary condition at $\omega=\omega_0$ and initial
conditions $p(\omega_i,t_i | \omega_0,t_i)=\delta(\omega_i-\omega_0)$ ($\delta(\cdot)$ stands
for the Dirac delta function). Here $p(\omega_i,t | \omega_0,t_i)$
is the probability that node $i$ has wealth $\omega_i$ at time $t$
given that it had $\omega_0$ at time $t_i$. The choice of a
reflecting boundary condition at $\omega=\omega_0$ is equivalent
to assume that $\beta$ is the overall growth rate of the number of nodes, that is, the composition of the birth and dead processes ruling the evolution of the number
of nodes.

Finding the solution for this problem is not an easy task.
Fortunately, we can take advantage of the fact that, when $\alpha
> \beta$, the average number of users of each node increases exponentially
and, since $D(\omega_i,t) = {\cal O} \left( A(\omega_i,t)\right)$,
fluctuations vanishes in the long time limit. Under this zero noise approximation, the number of hosts connected to an AS introduced
at time $t_i$ is
\begin{equation}
\omega_i(t|t_i)=\frac{\beta}{\alpha}\omega_0+(1-\frac{\beta}{\alpha})\omega_0
e^{\alpha(t-t_i)}.
\end{equation}
The probability density function of $\omega$ can be calculated in the long time limit as
\begin{equation}
p(\omega,t)=\beta e^{-\beta t} \int_0^t e^{\beta t_i} \delta(\omega-\omega_i(t|t_i)) dt_i
\end{equation}
which leads to
\begin{equation}
p(\omega,t)=
\displaystyle{ \frac{\tau(1-\tau)^{\tau} \omega_0^{\tau}}{(\omega-\tau\omega_0)^{1+\tau}}
\Theta(\omega_c(t)-\omega)},
\label{p_omega}
\end{equation}
where we have defined $\tau\equiv \beta/\alpha$ and the cut-off is
given by $\omega_c(t) \sim (1-\tau)\omega_0 e^{\alpha t} \sim
W(t)$. Thus, in the long time limit, $p(\omega,t)$
approaches a stationary distribution with an increasing
cut-off. In the case of the Internet, $\alpha \gtrsim \beta$ which implies an exponent smaller but close to 2. A similar result was obtained in \cite{Fayed}.

The key point in what follows is how to relate the number of users attached to an AS
with its degree. Our basic assumption is that vertices are
continuously adapting their bandwidth to the number of users they
have. However, once an AS decides to increase its bandwidth it
has to find a peer who, at the same time, wants to increase its
bandwidth as well. The reason is that connection costs among
ASs must be assumed by both peers. This fact differs from other
growing models in which vertices do not ask target vertices if they really want  to form those connections. Our model is, then, to be though of as a coupling
between a competition process for resources and adaptation of
vertices to their current situation, with the constraint
that connections are only formed between active nodes. Let $b_i(t|t_i)$ be the total bandwidth of an AS at time $t$ given that it was introduced at
time $t_i$. This quantity can include single connections with other ASs, {\it i. e.} the topological degree $k$, but it also accounts for connections which have higher capacity. This is equivalent
to say that the network is, in fact, weighted and $b_i$ is the weighted degree. To simplify the
model we consider that bandwidth is discretized in such a way that
single connections with high capacity are equivalent to multiple
connections between the same ASs.  Then, when a pair of ASs
agrees to increase their mutual connectivity the connection is newly formed if they were not
previously connected or, if they were, their mutual bandwidth increases by one unit. Now, we assume
that, at time $t$, each AS adapts its total bandwidth proportionally
to its number of users. We can write
\begin{equation}
b_i(t|t_i)=1+a(t)\left( \omega_i(t|t_i)-\omega_0 \right).
\label{bandwidth}
\end{equation}
Summing Eq.~(\ref{bandwidth}) for all nodes we get
\begin{equation}
a(t)=\frac{2B(t)-N(t)}{W(t)-\omega_0N(t)} \sim \frac{2B(t)}{W(t)},
\end{equation}
where $B(t)$ is the total bandwidth of the network which is, obviously, an upper bound to the total number of edges
of the network. This suggests that $B(t)$ will grow according to
$B(t)=B_0 e^{\delta' t}$, where $\delta' \gtrsim \delta$. Using this
assumption, we can express the individual bandwidth as $b_i(t|t_i) \sim
N(t)^{(\delta'-\alpha)/\beta}\omega_i(t|t_i)$. From this equation,
the scaling of the maximum bandwidth with the system size reads
$b_c(t)\sim N(t)^{\delta'/\beta}$, that is, faster than $N(t)$.
This implies that the network must necessarly contain
multiple connections. Then, we propose that degree and bandwidth are related, in a statistical
sense, through the following scaling relation
\begin{equation}
k(t|t_i) \sim \left[b(t|t_i) \right]^{\mu},
\label{degree_wealth}
\end{equation}
where the scaling exponent, $\mu<1$, is obtained by imposing that
the maximum degree scales linearly with $N(t)$ \footnote{Empirical
measurements made in \cite{Goh} showed such linear scaling in the
AS with the largest degree.}. This sets the scaling exponent to
$\mu=\beta/\delta'$. All four growth rates in the model are not independent but can be related by exploring the interplay between bandwidth, connectivity and traffic of the network. As the number of users grow, the global traffic of the Internet also grows, which means that ASs do not
only adapt their bandwidth to their number of users but to the
global traffic of the network. Therefore, $a(t)$ must be an increasing
function of $t$, which, in turn, implies that $\delta'>\alpha$.
Using this condition and summing Eq.~(\ref{degree_wealth}) for all
vertices, the scaling of the total number of connections is
$E(t)\sim N(t)^{2-\alpha/\delta'}$, which leads to $\delta'=\alpha \beta
/(2\beta-\delta)$. Combining this relation with Eqs.~(\ref{p_omega}), (\ref{bandwidth}) and (\ref{degree_wealth}), the
degree distribution reads
\begin{equation}
P(k)\sim \frac{\tau(1-\tau)^{\tau} \left[ \omega_0
a(t)\right]^{\tau}}{\mu} \frac{1}{k^{\gamma}} \Theta(k_c(t)-k),
\label{p_k}
\end{equation}
where the exponent $\gamma$ takes the value
\begin{equation}
\gamma=1+\frac{\beta}{2\beta-\delta}.
\end{equation}
Strikingly, the exponent $\gamma$ has lost any dependence on
$\alpha$ becoming a function of the growth rate of both the number
of ASs and the number of connections of the network. Using the empirical values
for $\beta$ and $\delta$, the predicted exponent is $\gamma =2.2 \pm 0.1$, in excellent agreement with the
values reported in the literature \cite{Falou99,Alexei,Alexei02}.

\begin{figure}
\epsfig{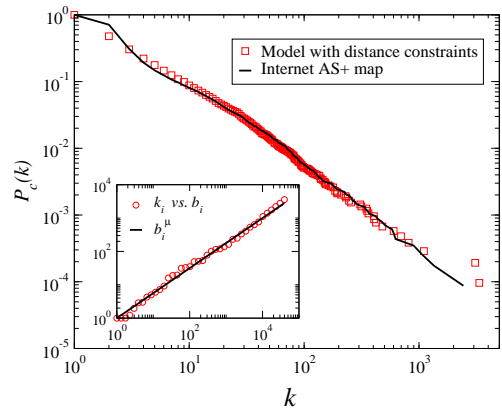} \caption{Cumulative degree
distribution ($P_c(k)=\sum_k P(k')$) for the extended AS map
compared to simulations of the model, $r=0.8$. Inset: AS's degree as a function of AS's bandwidth. The solid line stands for the scaling relation Eq.~(\ref{degree_wealth}) with $\mu=\beta/\delta'=0.75$.}
 \label{p_c}
\end{figure}

So far, we have been mainly interested
in the degree distribution of the AS map but not in the specific
way in which the network is formed. To fill this gap we have
performed numerical simulations that generate network topologies in nice agreement with real
measures of the Internet that go beyond the degree distribution. We consider a realistic geographical deployment of ASs and physical distance among them to take into account connection costs \cite{Yook}. Our algorithm, following the lines of the model, works in four steps:
\begin{enumerate}
\item At iteration t, $\Delta W(t)=\omega_0 N_0 (e^{\alpha
t}-e^{\alpha(t-1)})$ users join the network and choose provider
among the existing nodes using the preference rule Eq.~(\ref{preferential}).
\item $\Delta N(t)=N_0(e^{\beta t}-e^{\beta(t-1)})$ new ASs are
introduced with $\omega_0$ users each, those being randomly
withdrawn from already existing ASs. Newly created ASs are located in a two dimensional plane following a fractal set of dimension $D_f=1.5$ \cite{Yook}.
\item Each AS evaluate its
increase of bandwidth, $\Delta b_i(t|t_i)$, according to Eq.~(\ref{bandwidth}). 
\item A pair of nodes, $(i,j)$, is chosen with
probability proportional to $\Delta b_i(t|t_i)$ and $\Delta
b_j(t|t_j)$ respectively, and, whenever they both need to increase their
bandwidth, they form a connection with probability $D(d_{ij},\omega_i,\omega_j)$. This function takes into consideration that, due to connection costs, physical links over long distances are unlikely to be created by small peers. Once the first connection has been formed, they create a new connection with probability $r$, whenever they still need to increase their bandwidth. This
step is repeated until all nodes have the desired bandwidth.
\end{enumerate}
It is important to stress the fact that nodes must be chosen with
probability proportional to their increase in bandwidth at each
step. The reason is that those nodes that need a high bandwidth
increase will be more active when looking for partners to whom
form connections. Another important point is the role of the
parameter $r$. This parameter takes into account the balance
between the costs of forming connections with new peers and the
need for diversification in the number of partners. The effect of
$r$ in the network topology is to tune the average degree and the
clustering coefficient by creating more multiple connections. The
exponent of the degree distribution is unaffected except in the
limiting case $r \rightarrow 1$. In this situation, big peers will create a huge amount of multiple connections among them, reducing, thus, the maximum degree of the network. 
Finally, we chose an exponential form for the distance probability function $D(d_{ij},\omega_i,\omega_j) =e ^ {-d_{ij}/d_c(\omega_i,\omega_j)}$, where $d_c(\omega_i,\omega_j)= \omega_i\omega_j/\kappa W(t)$ and $\kappa$ is a cost function of number of users per unit distance, depending on the maximum distance of the fractal set. All simulations are performed using $\omega_0=5000$, $N_0=2$, $B_0=1$, $\alpha=0,035$, $\beta=0,03$, and $\delta'=0,04$, and the final size of the networks is $N \sim 11000$. Simulations will be compared to the AS+ extended map recorded on May 2001, as reported in \cite{Qian} that offers a better picture of the actual map.

\begin{figure}
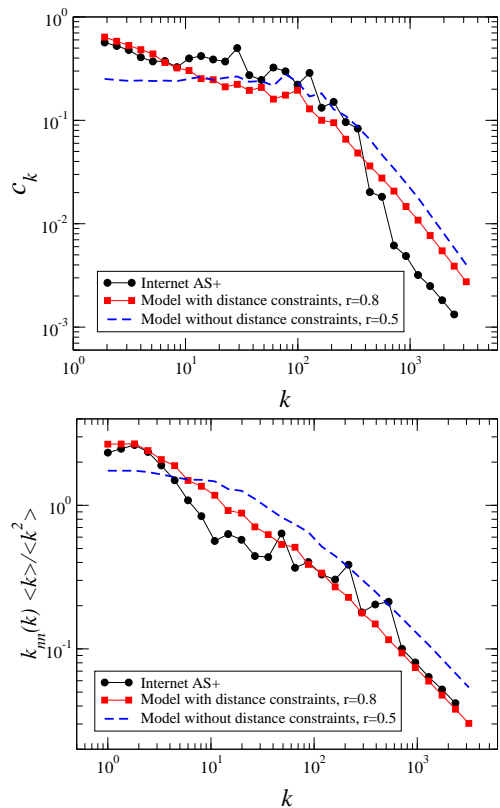

\begin{tabular}{c}
\epsfig{file=c_k.eps, width=6.5cm} \\
\epsfig{file=knn.eps, width=6.5cm} 
\end{tabular}
\caption{Clustering coefficient, $c_k$ (top) and normalised average nearest
neighbors degree, $\bar{k}_{nn}(k) \langle k \rangle / \langle k^2 \rangle$ (bottom), as functions of the node's degree for the extended autonomous system map (circles) and for the model with and without distance constraints (red squares and dashed line, respectively).}
 \label{c_k-knn}
\end{figure}

Fig.~\ref{p_c} shows simulation results for the cumulative degree distribution, in nice agreement to that measured for the AS+ map. The inset exhibits simulation results of the AS's degree as a function of the AS's bandwidth, conforming to the scaling relation in Eq.~(\ref{degree_wealth}). Clustering coefficient and average nearest neighbors degree are showed in Fig.~\ref{c_k-knn}. Dashed lines result from the model without distance constraints, whereas squares correspond to the model with distance constraints. Interestingly, the high level of clustering coming out from the model arises as a consequence of the pattern followed to attach nodes, so that only those AS willing for new connections will link. As can be observed in the figures, distance constraints introduce a disassortative component by inhibiting connections between small ASs so that the hierarchical structure of the real network is better reproduced. 

We conclude by pointing out that this work is a first attempt
towards a more realistic and complete modeling of the Internet,
which, for instance, is of utmost importance in new communication protocols
testing, which can be very sensitive to topological details. We would like to stress that the relevance of our model resides in the robustness of a simple statistical physics approach and, as a result, the 
unprecedented completedness of the topological description of the Internet and the novel 
insights into the dynamical processes leading to network formation.

\begin{acknowledgments}
We acknowledge R. Pastor-Satorras and A. Arenas for valuable comments and suggestions. This work  has been partially supported by DGES of the Spanish government, Grant No. BFM2003-08258, and EC-FET Open
project COSIN IST-2001-33555. M. B.  acknowledges financial support from the MCyT (Spain).
\end{acknowledgments}

\end{document}